\documentclass[10pt]{revtex4}
\newcount\driver
\newcount\bozza

\font\ottorm=cmr8\font\ottoi=cmmi8\font\ottosy=cmsy8%
\font\ottocss=cmcsc8%
\font\sixrm=cmr6\font\sixi=cmmi6\font\sixsy=cmsy6%
\font\fiverm=cmr5\font\fivesy=cmsy5
\font\fivei=cmmi5
\font\tenmib=cmmib10
\font\sevenmib=cmmib10 scaled 800

 2

\font\cs=cmcsc10
\font\sc=cmcsc10

\font\elevenrm=cmr11
\font\twelverm=cmr12
\font\ottorm=cmr8
\font\msytw=msbm9 scaled\magstep1

\font\msytwww=msbm5 scaled\magstep1
\font\indbf=cmbx10 scaled\magstep2

\font\ottorm=cmr8\font\ottoi=cmmi8\font\ottosy=cmsy8%
\font\ottocss=cmcsc8%
\font\sixrm=cmr6\font\sixi=cmmi6\font\sixsy=cmsy6%
\font\fiverm=cmr5\font\fivesy=cmsy5
\font\fivei=cmmi5

\def\ottopunti{\def\rm{\fam0\ottorm}%
\textfont0=\ottorm\scriptfont0=\sixrm\scriptscriptfont0=\fiverm%
\textfont1=\ottoi\scriptfont1=\sixi\scriptscriptfont1=\fivei%
\textfont2=\ottosy\scriptfont2=\sixsy\scriptscriptfont2=\fivesy%
\textfont4=\ottocss\scriptfont4=\sc\scriptscriptfont4=\sc%
\scriptfont4=\ottocss\scriptscriptfont4=\ottocss%
\textfont5=\tenmib\scriptfont5=\sevenmib\scriptscriptfont5=\fivei
\setbox\strutbox=\hbox{\vrule height7pt depth2pt width0pt}%
\normalbaselineskip=9pt\let\sc=\sixrm\normalbaselines\rm}

\mathchardef\BDpr = "0540  
\mathchardef\Bg   = "050D  

{\count255=\time\divide\count255 by 60 \xdef\hourmin{\number\count255}
        \multiply\count255 by-60\advance\count255 by\time
   \xdef\hourmin{\hourmin:\ifnum\count255<10 0\fi\the\count255}}

\def\openone{\leavevmode\hbox{\elevenrm 1\kern-3.63pt\twelverm1}}%
\def\*{\vglue0.5truecm}

\let\a=\alpha \let\b=\beta  \let\g=\gamma   \let\e=\varepsilon
      \let\k=\kappa 
\let\m=\mu    \let\n=\nu         \let\p=\pi    \let\r=\rho
 \let\t=\tau   \let\f=\varphi \let\c=\chi
   \let\o=\omega
 \let\D=\Delta  \let\L=\Lambda 
         
\let\O=\Omega 

\def\\{\hfill\break} \let\==\equiv

\let\io=\infty 

\let\0=\noindent

\def\media#1{{\langle#1\rangle}}

\def\tende#1{\,\vtop{\ialign{##\crcr\rightarrowfill\crcr
 \noalign{\kern-1pt\nointerlineskip} \hskip3.pt${\scriptstyle
 #1}$\hskip3.pt\crcr}}\,}
\def\circage{\lower2pt\hbox{$\,\buildrel > \over {\scriptstyle \sim}\,$}}
\def\otto{\,{\kern-1.truept\leftarrow\kern-5.truept\to\kern-1.truept}\,}

\def\T#1{{#1_{\kern-3pt\lower7pt\hbox{$\widetilde{}$}}\kern3pt}}
\def\VVV#1{{\underline #1}_{\kern-3pt
\lower7pt\hbox{$\widetilde{}$}}\kern3pt\,}
\def\W#1{#1_{\kern-3pt\lower7.5pt\hbox{$\widetilde{}$}}\kern2pt\,}

\def\indica{\leaders \hbox to 0.5cm{\hss.\hss}\hfill}
\def\guida{\leaders\hbox to 1em{\hss.\hss}\hfill}

   \def\qq{{\bf q}}
   \def\pp{{\bf p}}
 \def\xx{{\bf x}} \def\yy{{\bf y}} \def\zz{{\bf z}}
\def\hhh{{\bf h}}
\def\kk{{\bf k}}

\mathchardef\aa   = "050B
\mathchardef\bb   = "050C
\mathchardef\ggg  = "050D
\mathchardef\xxx  = "0518
\mathchardef\zzzzz= "0510
\mathchardef\oo   = "0521
\mathchardef\lll  = "0515
\mathchardef\mm   = "0516
\mathchardef\Dp   = "0540
\mathchardef\H    = "0548
\mathchardef\FFF  = "0546
\mathchardef\ppp  = "0570
\mathchardef\Bn   = "0517
\mathchardef\pps  = "0520
\mathchardef\fff  = "0527
\mathchardef\FFF  = "0508
\mathchardef\nnnnn= "056E

\def\to{\rightarrow}

\def\qed{\ \raise1pt\hbox{\vrule height5pt width5pt depth0pt}}

\def\indic{\hbox{\raise-2pt \hbox{\indbf 1}}}

\def\RRR{\hbox{\msytw R}} 
\def\rrr{\hbox{\msytwww R}} \def\CCC{\hbox{\msytw C}}
 
\def\NNN{\hbox{\msytw N}} 
 \def\ZZZ{\hbox{\msytw Z}}
 \def\zzz{\hbox{\msytwww Z}}

\def\V0{{\bf 0}}

\font\tenmib=cmmib10 
\font\sevenmib=cmmib7\font\fivemib=cmmib5 

\font\fivei=cmmi5\font\sixi=cmmi6\font\ottoi=cmmi8
\font\ottorm=cmr8\font\fiverm=cmr5\font\sixrm=cmr6
\font\ottosy=cmsy8\font\sixsy=cmsy6\font\fivesy=cmsy5
\font\ottocss=cmcsc8%

\mathchardef\Ba   = "050B  
\mathchardef\Bb   = "050C  
\mathchardef\Bg   = "050D  
\mathchardef\Bd   = "050E  
\mathchardef\Be   = "0522  
\mathchardef\Bee  = "050F  
\mathchardef\Bz   = "0510  
\mathchardef\Bh   = "0511  
\mathchardef\Bthh = "0512  
\mathchardef\Bth  = "0523  
\mathchardef\Bi   = "0513  
\mathchardef\Bk   = "0514  
\mathchardef\Bl   = "0515  
\mathchardef\Bm   = "0516  
\mathchardef\Bn   = "0517  
\mathchardef\Bx   = "0518  
\mathchardef\Bom  = "0530  
\mathchardef\Bp   = "0519  
\mathchardef\Br   = "0525  
\mathchardef\Bro  = "051A  
\mathchardef\Bs   = "051B  
\mathchardef\Bsi  = "0526  
\mathchardef\Bt   = "051C  
\mathchardef\Bu   = "051D  
\mathchardef\Bf   = "0527  
\mathchardef\Bff  = "051E  
\mathchardef\Bch  = "051F  
\mathchardef\Bps  = "0520  
\mathchardef\Bo   = "0521  
\mathchardef\Bome = "0524  
\mathchardef\BG   = "0500  
\mathchardef\BD   = "0501  
\mathchardef\BTh  = "0502  
\mathchardef\BL   = "0503  
\mathchardef\BX   = "0504  
\mathchardef\BP   = "0505  
\mathchardef\BS   = "0506  
\mathchardef\BU   = "0507  
\mathchardef\BF   = "0508  
\mathchardef\BPs  = "0509  
\mathchardef\BO   = "050A  
\mathchardef\BDpr = "0540  
\mathchardef\Bstl = "053F  

\def\V#1{{\bf#1}}
\let\aa=\Ba\let\fff=\Bf
\let\oo=\Bo
\let\nn=\Bn
\let\pps=\Bps\def\hhh={\V h}
\let\bb=\Bb

\def\RRR{\hbox{\msytw R}} 
\def\rrr{\hbox{\msytwww R}} \def\CCC{\hbox{\msytw C}}
 
\def\NNN{\hbox{\msytw N}} 
 \def\ZZZ{\hbox{\msytw Z}}
 \def\zzz{\hbox{\msytwww Z}}

\def\ins#1#2#3{\vbox to0pt{\kern-#2 \hbox{\kern#1 #3}\vss}\nointerlineskip}
\newdimen\xshift \newdimen\xwidth \newdimen\yshift
\newcount\griglia

\def\insertplot#1#2#3#4#5#6{%
\begin{figure}[h]
\begin{center}
\vspace{#2pt}
\begin{minipage}{#1pt}
#3
\ifnum\driver=1
\griglia=#6
\ifnum\griglia=1
\openout13=griglia.ps
\write13{gsave .2 setlinewidth}
\write13{0 10 #1 {dup 0 moveto #2 lineto } for}
\write13{0 10 #2 {dup 0 exch moveto #1 exch lineto } for}
\write13{stroke}
\write13{.5 setlinewidth}
\write13{0 50 #1 {dup 0 moveto #2 lineto } for}
\write13{0 50 #2 {dup 0 exch moveto #1 exch lineto } for}
\write13{stroke grestore}
\closeout13
\includegraphics{griglia.ps}\fi
\includegraphics{#4.ps}\fi
\ifnum\driver=2
\fi
\end{minipage}
\end{center}
\caption{#5}
\end{figure}
}

\newdimen\shift \shift=5truecm
\def\lb#1{%
\ifnum\bozza=1
\label{#1}\rlap{\kern\shift{\hfill$\scriptstyle#1$}}
\else\label{#1}
\fi}

\def\be{\begin{equation}}
\def\ee{\end{equation}}
\def\bea{\begin{eqnarray}}\def\eea{\end{eqnarray}}
\def\bean{\begin{eqnarray*}}\def\eean{\end{eqnarray*}}
\def\bfr{\begin{flushright}}\def\efr{\end{flushright}}
\def\bc{\begin{center}}\def\ec{\end{center}}
\def\ba#1{\begin{array}{#1}} \def\ea{\end{array}}
\def\bd{\begin{description}}\def\ed{\end{description}}
\def\Halmos{\hfill\vrule height10pt width4pt depth2pt \par\hbox to \hsize{}}

\def\const{{\rm const\,}}
\def\hw{{\hat w}}
\def\cA{{\cal A}}
\def\cB{{\cal B}}

\newtheorem{lemma}{Lemma}[section]
\newtheorem{theorem}{Theorem}[section]
\renewcommand{\theequation}{\arabic{section}.\arabic{equation}}

\newdimen\xshift \newdimen\xwidth \newdimen\yshift \newdimen\ywidth

\def\ins#1#2#3{\vbox to0pt{\kern-#2\hbox{\kern#1 #3}\vss}\nointerlineskip}

\def\eqfig#1#2#3#4#5{
\par\xwidth=#1 \xshift=\hsize \advance\xshift
by-\xwidth \divide\xshift by 2
\yshift=#2 \divide\yshift by 2
\line{\hglue\xshift \vbox to #2{\vfil
#3 \includegraphics{#4.ps}
}\hfill\raise\yshift\hbox{#5}}}

\def\8{\write12}

\begin{document}

\title{The ground state energy of the weakly interacting\\ Bose gas 
at high density\footnote{\copyright\,2008 by the authors.  This paper may be
  reproduced, in its entirety, for non-commercial purposes.\\ R.S.
acknowledges partial support by U.S. National Science
Foundation grant PHY-0652356.}} 

\author{Alessandro Giuliani}
\affiliation{Dipartimento di Matematica, Universit\`a degli Studi Roma Tre,
L.go S. L. Murialdo 1, 00146 Roma, Italy}
\author{Robert Seiringer}
\affiliation{Department of Physics, 
Princeton University, Princeton NJ 08544, USA}
\vspace{1cm}
\date{November 7, 2008}

\begin{abstract} 
  We prove the Lee-Huang-Yang formula for the ground state energy of
  the 3D Bose gas with repulsive interactions described by the
  exponential function, in a simultaneous limit of weak coupling and
  high density. In particular, we show that the Bogoliubov
  approximation is exact in an appropriate parameter regime, as far as
  the ground state energy is concerned.
\end{abstract}

\maketitle

\section{Introduction}

We consider a three dimensional system of $N$ interacting bosons
in a cubic (periodic) box $\L$ of side length $L$, described 
by the Hamiltonian: 
\be H_N=-\sum_{i=1}^N\D_i+\frac{a_0}{R_0^3}\sum_{1\le i<j\le N}v_{R_0}
(\xx_i-\xx_j)\;.\lb{1.1}\ee
Here $\xx_i\in\L$, $i=1,\ldots,N$, are the positions of the particles,
and $\D_i$ denotes the Laplacian with respect to $\xx_i$. Units are
chosen such that $\hbar=2m=1$, where $m$ is the mass of the
particles. The interaction potential is taken to be $v_{R_0}(\xx)=
\sum_{{\bf n}\in\zzz^3}e^{-|\xx+{\bf n}L|/R_0}$, and $a_0$ and $R_0$
are positive constants. The Hamiltonian (\ref{1.1}) operates on
symmetric wave functions in the Hilbert space $L^2(\L^N,d\xx_1\cdots
d\xx_N)$, as is appropriate for bosons.

We are interested in the ground state energy $E_0(N)$ of (\ref{1.1})
in the thermodynamic limit when $N$ and $|\L|$ tend to infinity with the 
density $\r=N/|\L|$ fixed, and in a weak coupling regime 
$a_0\ll \min\{\r^{-1/3},R_0\}$. 
The constant $a_0$ is the first Born approximation to the scattering
length $a$ of the potential $(a_0/R_0^3)e^{-|\xx|/R_0}$, which is 
defined as usual as $a=\lim_{|\xx|\to\io}|\xx|(1-\psi_0(\xx))$, with $\psi_0$ a solution
to the zero energy scattering 
equation 
\be -2\D\psi(\xx) +\frac{a_0}{R_0^3}e^{-|\xx|/R_0}\psi(\xx) =0\ee
with boundary condition $\lim_{|\xx|\to\infty} \psi(\xx)=1$. It is well known that, if 
$a_0/R_0\ll1$, $a/a_0$ can be written in terms of a convergent series 
in powers of $a_0/R_0$ (Born series), which will be denoted by 
$a=a_0+\sum_{k\ge 1} a_k$, and whose first non-trivial term is given by 
\be a_1=-\frac1{128\p^3}\int_{\rrr^3}d\kk \, \frac{\n(\kk)^2}{\kk^2}= - \frac{5\pi}{16} \frac{a_0^2}{R_0}\;, \ee
where
\be\n(\kk)=\frac{a_0}{R_0^3}\int_{\rrr^3} d\xx\; e^{-|\xx|/R_0}
e^{-i\kk\xx}=\frac{8\p a_0}{\big[1+(\kk R_0)^2\big]^2}\;.\lb{1.3}\ee

The current understanding of the properties of the ground state of (\ref{1.1}) 
is based
on the pioneering work of Bogoliubov \cite{Bo}, who developed an approximate
theory of the ground state of weakly repulsive bosons. 
In the regime  
$1\gg a/R_0\gg \sqrt{\r a^3}\gg 
(a/R_0)^2$, Bogoliubov's theory predicts \cite{LSSY} that the ground state 
energy $E_0(N)$ of (\ref{1.1}) in the thermodynamic limit $N,|\L|\to\io$, 
with $\r=N/|\L|$ fixed, satisfies 
\be \lim_{N,|\L|\to\io}\frac{E_0(N)}
{N}= 4 \p\r a \left(   1+\frac{128}{15\sqrt\p}\sqrt{\r a^3}
+o\big(\sqrt{\r a^3}\big) \right) \;.\lb{1.2}\ee
This formula was first derived in \cite{LHY} and it is known as the 
Lee-Huang-Yang formula. 
Our goal is to prove that the expression (\ref{1.2}) is asymptotically correct 
in a regime such that $a\ll\r^{-1/3}\ll R_0$, that is a {\it weak coupling}
and {\it high density} regime. We shall prove the following theorem. \\

{\bf Theorem 1.} {\it  Let $Y =\r a^3$. 
There exists a positive constant $d_0$, which can be 
chosen to be $d_0=1/69$, such that, if
$0<d< d_0$ and $a/R_0 =O(Y^{1/2-d})$, then 
(\ref{1.2}) is valid, asymptotically as $Y\to 0$.}\\

This result represents the first rigorous proof of the Lee-Huang-Yang formula for 
the ground state energy of a weak-coupling Bose gas. We note that for $d<1/6$, 
$R_0^3\rho \gg 1$ and hence Theorem~1 concerns the high density regime.  Our 
result is not expected to be optimal. In fact, the formula (\ref{1.2}) is expected 
to hold even for $d=1/2$, i.e., for $a/R_0$ fixed and $\r a^3\to 0$ \cite{LHY}, 
but the Bogoliubov approximation is not valid in this case.  The prediction of 
Bogoliubov's theory is that (\ref{1.2}) should be valid for any $0<d<1/4$, i.e., 
in the regime $a/R_0\gg\sqrt{\r a^3}\gg (a/R_0)^2$. The latter condition is 
necessary in order that $a \approx a_0 + a_1$ to the desired accuracy (i.e., up to 
error terms that are much smaller than $a_0\rho \sqrt{\r
  a_0^3}$), and the former is certainly needed since $E_0(N)/N \leq
4\pi \rho a_0$ (i.e., the right side of (\ref{1.2}) must be equal
to $4\p\r a_0$ plus a {\it negative} correction, which requires
$|a_1|\gg a_0 \sqrt{\r a_0^3}$).

For simplicity, we shall restrict our attention to interaction
potentials given by the exponential function. Our proof can be adapted
to a larger class of potentials, including the Yukawa
potential. In our proof, however, we need the potential to be positive
definite, with a Fourier transform satisfying nice decay properties as
$|\kk|\to\io$ (e.g., polynomial decay), and our proof does not
immediately extend beyond this class.  Such restrictive condition is
not supposed to have any physical relevance, of course, and
(\ref{1.2}) should hold for much more general repulsive potentials. We
hope that the technical restrictions under which we proved Theorem~1
will be eliminated in future works \footnote{In the process of
  writing up this paper, we learned that E.H. Lieb and J.P. Solovej
  managed to prove the analogue of Theorem 1 for a larger class of
  repulsive potentials and in the larger regime $0<d<1/6 + \epsilon$
  for some $\epsilon>0$. We thank them for communicating their results
  to us.}.

\bigskip

The proof of Theorem~1 will proceed in two steps: we will get upper
and lower bounds with the correct asymptotic form.  The proof of the
upper bound is based on a computation of the variational energy
corresponding to the Bogoliubov trial wave function, following ideas
of Girardeau and Arnowitt \cite{GA}, see the next section.
 
The strategy of the proof of a lower bound will follow closely the one
of Lieb and Solovej in \cite{LS}, where the ground state energy of
bosonic jellium was investigated. We shall first localize the
Hamiltonian in boxes of size $\ell$. Using the positivity of the
Fourier transform of the exponential interaction, we shall derive a
preliminary estimate on the ground state energy and, correspondingly,
on the degree of condensation in the small boxes.  With this a priori
bound on the number of particles $n_+$ outside the condensate, we
shall be able to bound from below the full Hamiltonian by the
Bogoliubov Hamiltonian minus an error term, depending on the a priori
bound on $n_+$.  The key point is that it is possible to find a
scaling regime for $a_0$ and $R_0$ such that the new error term is
much smaller than the preliminary one, as $Y\to 0$.  With this 
improved bound on the ground state energy we shall obtain new
improved bounds on the size of fluctuations of $n_+$
that, in combination with the bounds for the ground
state energy, will allow us to conclude the desired lower bound.

\section{The upper bound}
\setcounter{equation}{0}

Let us first derive an upper bound to the ground state energy, asymptotically
agreeing with (\ref{1.2}). In second quantized form, the Hamiltonian $H_N$ 
can be rewritten as:
\be H_N=\sum_{\kk}\kk^2 c^{\dagger}_\kk c_\kk +\frac{1}{2|\L|}
\sum_{\kk,\qq,\pp}
\n(\pp) c^{\dagger}_{\kk+\pp}c^{\dagger}_{\qq-\pp}c_{\kk}c_{\qq}\;,
\label{3.1}\ee
where the sums run over vectors of the form $2\p\nn/L$, $\nn\in\ZZZ^3$,
and $c^{\dagger}_\kk, c_\kk$ are standard bosonic creation and annihilation 
operators, 
associated with the canonical basis of plane waves (for an introduction, see,
e.g., \cite{LSSY}). Following \cite{GA}, 
we choose the following variational state,
inspired by Bogoliubov's approximate treatment of the weak coupling
Bose gas:
\be |\O_{B,N}\rangle=\exp\Big\{\frac12
\sum_{\kk\neq\V0}\psi(\kk)\big( \b_\V0^{-1}\a_\kk-\b_\V0\a_\kk^{\dagger}\big)
\Big\}|\O_N\rangle\label{3.2}\ee
where:\\
1) $|\O_N\rangle=(N!)^{-1/2}(c_\V0^{\dagger})^N|0\rangle$ 
is the ground state for $N$ non-interacting particles;\\
2) the operator $\a_\kk$ is the pair annihilation operator $\a_\kk=c_\kk 
c_{-\kk}$;\\
3) if we denote by $N_\V0=c^{\dagger}_\V0 c_\V0$ the number operator associated
to the constant wave function, $\b_0$ is the partial isometry defined by
\be \b_{\bf 0}^{1/2}=
c_\V0 N_\V0^{-1/2}\;,\qquad\b_\V0^{-1/2}=(\b_0^{1/2})^{\dagger}
=N_\V0^{-1/2}c_\V0^\dagger \;,\label{3.3}\ee
having the properties 
\bea &&\b_\V0|\O_N\rangle=|\O_{N-2}\rangle\quad (N\ge 2)\;,\qquad 
\b_\V0^\dagger|\O_N\rangle=\b_\V0^{-1}|\O_N\rangle=|\O_{N+2}\rangle\;,\nonumber\\
&&[\b_\V0,N_\V0]=2\b_\V0\;,\qquad [\b_\V0^{-1},N_\V0]=-2\b_\V0^{-1}\;,\nonumber\\
&&[\b_\V0,c_\kk]=[\b_\V0,c_\kk^\dagger]=0\quad (\kk\neq\V0)\;;\label{3.4}\eea
4) $\psi$ is a continuous function from $\RRR^3$ to $\RRR$.

Note that $|\O_{B,N}\rangle$ is normalized, and that the particle
number is equal to $N$. The variational principle implies the upper
bound
\be E_0(N)\le \media{\O_{B,N}|H_N|\O_{B,N}}\;.\label{3.5}\ee
Following \cite{GA}, 
after a lengthy but straightforward computation, we find that
in the thermodynamic limit
\bea&& \lim_{N\to\io}
\frac1N\media{\O_{B,N}|H_N|\O_{B,N}}=\frac12\r\n(\V0)+
\r^{-1}\int\frac{d\kk}{(2\p)^3}\Big[\kk^2+\r_0\n(\kk)+
\frac12 I_2(\kk)\Big]
\sinh^2\psi(\kk)\nonumber\\
&&\hskip4.5truecm-\r^{-1}\int\frac{d\kk}{(2\p)^3}\Big[\r_0\n(\kk)-
\frac12 I_1(\kk)\Big]\sinh\psi(\kk)\cosh\psi(\kk)\;,\label{3.6}\eea
where:
\bea && \r_0=\r-\int\frac{d\qq}{(2\p)^3}\sinh^2\psi(\qq)\nonumber\\
&& I_1(\kk)=\int\frac{d\qq}{(2\p)^3}\n(\kk-\qq)\sinh\psi(\qq)\cosh\psi(\qq)
\nonumber\\
&& I_2(\kk)=\int\frac{d\qq}{(2\p)^3}\n(\kk-\qq)\sinh^2\psi(\qq)\;.
\label{3.7}\eea
Choosing $\psi(\kk)=\frac12\tanh^{-1} \frac{\r\n(\kk)}{\kk^2+\r\n(\kk)}$,
we find that 
\bea &&
\sinh^2\psi(\kk)=\frac12\frac{\kk^2+\r\n(\kk)-\sqrt{\kk^4+2\r\n(\kk)\kk^2}}
{\sqrt{\kk^4+2\r\n(\kk)\kk^2}}\;,\nonumber\\
&&
\sinh \psi(\kk)\cosh\psi(\kk)=\frac12\frac{\r\n(\kk)}
{\sqrt{\kk^4+2\r\n(\kk)\kk^2}}\;.\eea
Recall that $\n(\kk)$ is given in (\ref{1.3}), and that $a/R_0=O(Y^{1/2-d})$. A simple calculation shows that 
\be \r_0=\r\big(1+O(\sqrt{\r a^3})\big)\;\label{3.7a} \ee
for any $d>0$.  Moreover, we have the bounds
\be |I_1(\kk)|\le C\r a_0 \frac{a_0}{R_0}\frac1{\left[1+(\kk R_0)^2\right]^2
}\;,\qquad
|I_2(\kk)|\le C\r a_0 \sqrt{\r a_0^3}  \frac1{\left[1+(\kk R_0)^2\right]^2}\ee
for a suitable constant $C$.
Substituting these bounds into (\ref{3.6}) we find that 
\bea &&\frac1{4\p\r a_0}\lim_{N\to\io}
\frac1N\media{\O_{B,N}|H_N|\O_{B,N}}= \label{3.7b}\\
&&\hskip1.truecm = 1-\frac{1}{2\p\r^2a_0}
\int_{\rrr^3}\frac{d \kk}{(2\p)^3}\,\Big(
\kk^2+\r\n(\kk)-\sqrt{\kk^4+2\r\n(\kk)\kk^2}\Big)
+o(\sqrt{\r a^3})\nonumber\eea
for $0<d<1/4$. 
A computation \cite{LSSY} shows that for $d>0$ the integral on the right side equals 
\be \frac{a_1}{a_0}+\frac{128}{15\sqrt\p}\sqrt{\r a_0^3}+ 
o(\sqrt{\r a^3})\;.\label{3.7c}\ee
Noting that $a/(a_0+a_1) = O(a_0/R_0)^2 \ll Y^{1/2}$ for $d<1/4$ this yields the desired result.

\bigskip

{\bf Remark.} The upper bound we have just derived yields the desired
expression for any $0<d<1/4$. By suitably modifying the trial function
$\psi(\kk)$ above, one can actually show that the upper bound holds
for any $0<d<1/2$ \cite{ESY}. For $d=1/2$, however, the ansatz
(\ref{3.2}) can not be expected to yield the Lee-Huang-Yang formula,
even for the optimal choice of $\psi$.

\section{The lower bound}\label{lower}
\setcounter{equation}{0}

We shall split the lower bound into several parts. The strategy is
similar to the proof of the lower bound on the ground state energy of
jellium by Lieb and Solovej in \cite{LS}, and we shall refer to their
paper for several essential ingredients.

\subsection{Sliding and localizing}

We start by rewriting (\ref{1.1}) in the form
\bea H'_N=H_N-4\p N\r a_0=&&-\sum_{i=1}^N\D_i+\frac{a_0}{R_0^3}
\Bigl[\sum_{1\le i< j\le N}v_{R_0}(\xx_i-\xx_j)-\\\nonumber
&&-\r\sum_{i=1}^N\int_\L d\yy\, v_{R_0}(\xx_i-\yy)+
\frac{\r^2}{2}\int\!\!\!\int_{\L\times\L}\,d\xx\, d\yy
\,v_{R_0}(\xx-\yy)\Bigr]\;,\lb{2.1}\eea
with $\rho=N/|\L|$. 
We shall use the sliding method of \cite{CLY} to reduce the problem to a small box. 

Let $t$, with $0<t<1/2$, be a parameter which we shall choose later to
depend on $\rho$ in such a way that $t\to0$ as $\rho R_0^3\to\infty$.
Let $\chi\in C^\infty_0(\RRR^3)$ be supported in
$\left[(-1+t)/2,(1-t)/2\right]^3$, $0\leq \chi\leq1$, with $\chi(\xx)=1$
for $\xx$ in the smaller box $\left[(-1+2t)/2,(1-2t)/2\right]^3$, and
$\chi(\xx)=\chi(-\xx)$.  Assume that all $m$-th order derivatives of
$\chi$ are bounded by $C_m t^{-m}$, where the constants $C_m$ depend
only on $m$ and are, in particular, independent of $t$. If $M\in\NNN$
and $\ell=M^{-1}L$, let $\chi_\ell(\xx)$ be a function on the torus
$\L$ defined by $\c_\ell(\xx)=\sum_{{\bf
    n}\in\zzz^3}\chi\big(\ell^{-1}(\xx+{\bf n}L)\big)$. For given $\chi$
we also define $\gamma>0$ by $\gamma^{-1} =
\int\chi(\yy)^2\,d\yy$, and note that $1\leq \gamma\leq(1-2t)^{-3}$. 
We shall prove the following.

{\lemma\label{III.1} Let $\ell t R_0^{-1}$ be large enough.
There exists a function of the form $\omega(t)=\const t^{-1}R_0/\ell$  
such that if we set $R^{-1}=R_0^{-1}+\o(t)/\ell$ and
\be w_R^\L(\xx,\yy)=
\chi_\ell(\xx) v_{R}(\xx-\yy)\chi_\ell(\yy)\lb{2.2}\ee
then the potential energy satisfies 
\bea \lefteqn{\sum_{1\le i< j\le N}v_{R_0}(\xx_i-\xx_j)
-\r\sum_{i=1}^N\int_\L d\yy\, v_{R_0}(\xx_i-\yy)+
\frac{\r^2}{2}\int\!\!\!\int_{\L\times\L}\,d\xx\, d\yy
\,v_{R_0}(\xx-\yy)\ge}&&\\
&&\displaystyle\geq\frac{\gamma R}{R_0}\sum_{{\bf m}\in[1,\ldots,M]^3
}\ \int_{Q_{\bf m}}\ \frac{d\zz}{\ell^3}\ \Bigl\{
\sum_{1\leq i<j\leq N}w_R^\L\left(\xx_i+\zz,\xx_j+\zz\right)
-\rho\sum_{j=1}^N  \int_{\L}d\yy\,
w_R^\L\left(\xx_j+\zz,\yy+\zz\right)+\nonumber\\
&&{}+\frac{1}{2}\rho^2
\int\!\!\!\int_{\L\times\L}\,d\xx\,d\yy\,
w_R^\L\left(\xx+\zz,\yy+\zz\right)\Bigr\}-N\frac{\omega(t) R}{2\ell}-
\const N^2e^{-L/(2R)}\;,\lb{2.3}  \nonumber\eea
where $Q_{\bf m}$ is a cube of side length $\ell$ and centered at ${\bf m}\ell$ 
(so that the collection $\{Q_{\bf m}\}_{{\bf m}\in[1,\ldots,M]^3}$
paves the torus $\L$).}\\

The proof of Lemma~\ref{III.1} utilizes the following lemma, whose
proof will be given after the proof of Lemma~\ref{III.1}. An analogous
result for the Yukawa interaction potential was proved in
\cite[Lemma~2.1]{CLY}.

{\lemma\label{III.2a} Let $K: \RRR^3 \to \RRR $ be given by
\be K(\zz) = e^{-\nu |\zz|}\left(1 - \frac{e^{-\omega |\zz|}}{1+\omega/\nu} 
h(\zz)\right)\lb{A.1}
\ee
with $\nu\ge\omega > 0$. Let $h$ satisfy {\rm (i)} $h$ is a
$C^6$ function of compact support; {\rm (ii)} $h(0) =1$; {\rm (iii)}
all its $m$-th order derivatives, $1\le m\le 6$, are bounded by $C
t^{1-m}$ for some constants $C>0$ and $t>0$.  
Assume further that $h(\zz)=h(-\zz)$ so
that $K$ has a real Fourier transform. There exists a constant $C_1$ 
(depending only on $C$ but not on $t$, $\omega$ or $\nu$) such that,
if $\min\{1,\o\}\n t\ge C_1$, then $K$ has a positive Fourier transform.}
\\

{\sc Proof of Lemma~\ref{III.1}.} We calculate
\bea && \frac{\g R}{R_0}\sum_{{\bf m}\in[1,\ldots,M]^3
}\ \int_{Q_{\bf m}}\ \frac{d\zz}{\ell^3}\ 
w_R^\L\left(\xx+\zz,\yy+\zz\right)=\lb{2.4}\\
&&\qquad =\frac{\g R}{R_0}\int_\L\,\frac{d\zz}{\ell^3}\,
\chi_\ell(\xx+\zz)\,v_{R}(\xx-\yy)\,\chi_\ell(\yy+\zz)\,=\,\frac{R}{R_0}
h_\ell(\xx-\yy)\,v_{R}(\xx-\yy)\;,\nonumber \eea 
where we have set $h_\ell=\gamma\,\ell^{-3}\,\chi_\ell*\chi_\ell$.
Note that $h_\ell(\xx)$ vanishes if $\|\xx\|\ge \ell$, so we can naturally
introduce a function $h:\RRR^3\to\RRR$ of compact support and vanishing 
outside the cube of side 2 centered at the origin, such that
$h_\ell(\xx)=\sum_{{\bf n}\in\zzz^3}h\big(\ell^{-1}(\xx+{\bf n}L)\big)$. 
Note that: (i) $h({\bf 0})=1$; (ii) $h$ has a quadratic maximum at 
$\zz={\bf 0}$; (iii) $h$ is an even $C^\io$ function of compact support; 
(iv) all $m$-th order derivatives of $h$, $m\ge 1$,
are bounded by $C_m t^{1-m}$, where the constants $C_m$ depend only on $m$ 
and are, in particular, independent of $t$. The function $h$ thus satisfies
all the hypothesis of Lemma~\ref{III.2a}. Note also that 
the role of $\n$ and $\o$ in Lemma~\ref{III.2a} are here played by $\ell R_0^{-1}$ and 
$\o(t)$ respectively. So, if $\o(t)\ge C_1 R_0\ell^{-1}t^{-1}$, 
where $C_1$ is the 
constant appearing in the statement of Lemma~\ref{III.2a}, we then conclude from it that
the Fourier transform of the 
function $K(\xx)=e^{-|\xx|/R_0}-h(\ell^{-1}\xx)e^{-|\xx|/R}(R/R_0)$,
is positive. Now, defining $K^\L(\xx)=
\sum_{{\bf n}\in\zzz^3}K(\xx+{\bf n} L)$ and $\varphi(\xx)=v_{R_0}(\xx)-
h_\ell(\xx)v_R(\xx)(R/R_0)$, we note that 
$\varphi(\xx)=K^\L(\xx)+ R(\xx)$, with $|R(\xx)|\le \const
e^{-(L-\ell)/R}$. 
Because of positivity of the Fourier transform of $K$, 
\bea&& \sum_{1\leq i<j\leq N}\f(\xx_i-\xx_j) -\rho\sum_{j=1}^N
\int_{\L}\ \f(\xx_i-\yy)+\frac{1}{2}\rho^2\ \int\!\!\!\int_{\rrr^3\times\rrr^3}
\ \f(\xx-\yy)
\,d\xx\,d\yy \geq\nonumber\\
&& \hskip4.truecm
\geq -\frac{N}2 K^\L(\V0)-\const N^2 e^{-(L-\ell)/R}\;.\lb{2.5}\eea
Since $K^\L(\V0) \leq R\omega/\ell + \const \exp(-L/R_0)$ this implies (\ref{2.3}).\qed\\

{\sc Proof of Lemma~\ref{III.2a}.} We write $h(\zz) = 1 + q(\zz) + F(\zz)$, 
where $q(\zz)$ is an even polynomial of degree $4$ that vanishes at the origin,
and $F(\zz)\leq C t^{-5} |\zz|^6$. 
The Fourier transform of $e^{-\nu |\zz|} - e^{-(\nu+\omega)|\zz|}/(1+
\omega/\nu)$ is given by 
\be
\frac{8\pi \nu}{\left(\nu^2+\pp^2\right)^2} -
\frac{8\pi \nu}{\left((\nu+\omega)^2+\pp^2\right)^2} \geq \frac{ 48 
\pi \nu^2 \omega}{\left( (\nu+\omega)^2 +\pp^2\right)^3 }\,.
\ee
Moreover, the Fourier transform of $q(\zz) e^{-(\nu+\omega)|\zz|}$ is   
\be
q( i \nabla_\pp ) \frac{8\pi (\nu +\omega) }{\left( (\nu+\omega)^2 + 
\pp^2\right)^2 }
\ee
whose absolute value, if $\nu t \ge C_1$, can be bounded above by 
$\const\cdot \nu t^{-1} 
[(\nu+\omega)^2+\pp^2]^{-3}$ (here we used that $q$ is assumed to be 
even and that its $m$'th order coefficients are bounded by $C t^{1-m}$). 
Finally, we claim that the Fourier transform of 
$F(\zz) e^{-(\nu+\omega)|\zz|}$ is bounded by $\const\cdot \nu^{-3} t^{-5}  
[(\nu+\omega)^2 + \pp^2]^{-3}$. 
To see this, note that $F(\zz) e^{-(\nu+\omega)|\zz|}$ is a $C^6$ function, 
and hence 
\be
\left( (\nu+\omega)^2 + \pp^2\right)^3 \int F(\zz) e^{-(\nu+\omega) |\zz|} 
e^{-i \pp\cdot \zz} d\zz  = \int \left[ \left( (\nu+\omega)^2 - 
\Delta\right)^3 F(\zz) e^{-(\nu+\omega)|\zz|} \right] e^{-i \pp \cdot \zz} 
d\zz \,.
\ee
It is not difficult to see that the latter integral is bounded by $C t^{-5} 
\n^{-3}$. 
After collecting all the terms, we arrive at the statement of the lemma.\qed \\

Below, we shall choose the parameters $t$ and $\ell$ as functions of $\rho,R_0$
and $a_0$. We shall choose them in such a way that 
$t\ll 1$ and $\ell\gg R_0$. Moreover, we will have conditions of the form
\be\frac{\ell t}{R_0}\to \io\;,\qquad \frac{a_0}{R_0^2}\frac{\o(t)}{\ell}
\frac1{\r a_0\sqrt{\r a_0^3}}\to 
0\;,\qquad \text{and}\quad  \frac{R}{R_0} \to 1 \lb{2.6}\ee
as $\rho a_0^3\to 0$, such that the error in the specific ground state
energy corresponding to the term $N\omega(t)R/(2\ell)$ in (\ref{2.3})
is much smaller than $N\r a_0\sqrt{\r a_0^3}$, which is the precision
to which we want to compute the ground state energy.
 
Consider now the $n$-particle Hamiltonian 
\begin{equation}
H^{n}_{{\bf m},\zz}=-\sum_{j=1}^n\Delta^{(j)}_{Q_{{\bf m},\zz}}
+\frac{\g a_0 R}{R_0^4} W_{\zz},\lb{2.7}
\end{equation}
where we have introduced the Neumann Laplacian  
$\Delta^{(j)}_{Q_{{\bf m},\zz}}$ in the cube
$Q_{{\bf m},\zz}=Q_{\bf m}+\zz$ and the potential
\bea W_{\zz}(\xx_1,\ldots,\xx_n)&=&
\sum_{1\leq i<j\leq n}w_R^\L\left(\xx_i+\zz,\xx_j+\zz\right)
-\rho\sum_{j=1}^n  \int_{\L}\,d\yy\, w_R^\L\left(\xx_j+\zz,\yy+\zz\right)
\nonumber\\ &&+\frac{1}{2}\rho^2\int\!\!\!\int_{\L\times\L}\,d\xx\,d\yy\, 
w_R^\L\left(\xx+\zz,\yy+\zz\right)\,.\lb{2.8}\eea
Denoting by $E^n_{{\bf m},\zz}$ the ground state energy of the Hamiltonian
$H^{n}_{{\bf m},\zz}$ in (\ref{2.7}) considered as a bosonic Hamiltonian for $n$ particles confined to the cube $Q_{{\bf m,\zz}}$, and using Lemma \ref{III.1}, we find that 
the ground state energy $E_0$ of (\ref{1.1}) can be bounded below by
\be E_0\geq 4\p N\r a_0+\sum_{{\bf m}\in[1,\ldots,M]^3
}\ \int_{Q_{\bf m}}\ \frac{d\zz}{\ell^3}\,\inf_{1\le n\le N}
E^n_{{\bf m},\zz}-N\frac{a_0\omega(t) R}{2\ell R_0^3}-\const N^2 
\frac{a_0}{R_0^3}e^{-L/(2R)}. \lb{2.9}\ee
Note that all the Hamiltonians $H^{n}_{{\bf m},\zz}$ are unitarily equivalent 
to 
\be -\sum_{j=1}^n\Delta^{(j)}_{\ell} +\frac{\g a_0 R}{R_0^4} \Big[\sum_{1\leq i<j\leq n}w_R^\L
\left(\xx_i,\xx_j\right)
-\rho\sum_{j=1}^n  \int_{\L}\,d\yy\, w_R^\L\left(\xx_j,\yy\right)
+\frac{1}{2}\rho^2\int\!\!\!\int_{\L\times\L}\,d\xx\,d\yy\, 
w_R^\L\left(\xx,\yy\right)\Big]\lb{2.10} \ee
where $\Delta^{(j)}_{\ell}$ denotes the Neumann Laplacian for the $j$--th particle
in the cube $[-\ell/2,\ell/2]^3$. As a consequence, in the $L\to\io$ limit,
we have reduced the problem  to studying the Hamiltonians $H^n_\ell$ on $L^2([-\ell/2,\ell,2]^{3n})$, given by 
\bea&&H^n_\ell=-\sum_{j=1}^n\Delta^{(j)}_{\ell}\lb{2.10a}\\
&& + \frac{\g a_0 R}{R_0^4}  \Big[\sum_{1\leq i<j\leq n}w_R
\left(\xx_i,\xx_j\right)
-\rho\sum_{j=1}^n  \int_{\rrr^3}\,d\yy\, w_R\left(\xx_j,\yy\right)
+\frac{1}{2}\rho^2\int\!\!\!\int_{\rrr^3\times\rrr^3}\,d\xx\,d\yy\, 
w_R\left(\xx,\yy\right)\Big]\nonumber\eea
with $w_R(\xx,\yy)=\c(\xx\ell^{-1})e^{-|\xx-\yy|/R}\c(\yy\ell^{-1})$. If
$E^n_\ell$ is the ground state energy of $H^n_\ell$, 
from (\ref{2.9}) we infer that 
\be  \lim_{N\to\io}\frac{E_0}{N}\geq 4\p \r a_0+\frac1{\r\ell^3}
\,\inf_{n} E^n_\ell-\frac{a_0\omega(t) R}{2\ell R_0^3}. \lb{2.11}\ee

In the remainder of this paper we shall study the Hamiltonians (\ref{2.10a}).
For future reference, let us finally note that in second quantized 
form $H^n_{\ell}$ can be rewritten as 
\be H^n_{\ell}=\sum_{\pp} \pp^2a^{\dagger}_\pp a_\pp +
\frac{\g a_0 R}{R_0^4}  \Big[\frac12
\sum_{\pp\qq,\mm\Bn}\hw_{\pp\qq,\mm\Bn}a^{\dagger}_\pp 
a^{\dagger}_\qq a_\mm a_\Bn
-\rho\ell^3\sum_{\pp\qq}\hw_{\V0\pp,\V0\qq}a^{\dagger}_\pp a_\qq
+\frac12\rho^2\ell^6\hw_{\V0\V0,\V0\V0}\Big]\lb{2.18}\ee
where the sums over the momenta run over the values $\pp=(p_1,p_2,p_3)$
such that $\ell p_i/\p\in\ZZZ_+$, 
$a^{\dagger}_\pp,a_\pp$ are bosonic creation/annihilation operators 
corresponding to the orthonormal basis 
$\phi_\pp(\xx)=c_\pp\ell^{-3/2}\prod_{j=1}^3
\cos\big(p_j\p\ell^{-1}(x_j+\ell/2)\big)$, and the coefficients
$\hw_{\pp\qq,\mm\Bn}$ are defined as
\be \hw_{\pp\qq,\mm\Bn} = \int\!\!\!\int d\xx d\yy\, w_{R}(\xx,\yy)\,
\phi_\pp(\xx)\phi_\qq(\yy)\phi_\mm(\xx)\phi_\Bn(\yy)\;.\lb{2.19}\ee

\subsection{Scalings}\label{scalings}

Before proceeding with 
the proof of the lower bound, let us make a few
remarks on the choice of the parameters $a_0,R_0,\ell,t$.
We recall that our purpose is to compute the ground state energy
of (\ref{1.1}) up to terms of the order $N\r a_0\sqrt{\r a_0^3}$,
asymptotically as $Y=\r a_0^3\to 0$. In the following we shall choose 
$a_0/R_0\sim Y^{1/2-d}$, $a_0/\ell\sim Y^{b+1/2}$ and $t\sim Y^\t$, 
where $d,b$ and $\t$ are positive scaling exponents. [Here by $f\sim g$ we mean 
that $C^{-1}g\le f\le C g$, for some universal constant $C$.] We shall require 
$d<1/4$. Note that 
the conditions $b,d>0$ and $d<1/4$ imply in particular that
$a_0\ll R_0\ll \ell$ and $a_0/\ell\ll\sqrt{
\r a_0^3}$
(two conditions that are of course necessary to be able
to neglect finite size effects due to the boxes of size $\ell$)
and that $a_0/R_0\gg \sqrt{\r a_0^3}\gg (a_0/R_0)^2$ (a condition that is 
necessary for the Bogoliubov approximation to be valid, as explained in the Introduction). 
In order to be able to prove that the various error terms in our estimates  
are much smaller than 
$N\r a_0\sqrt{\r a_0^3}$ we will be forced to require that $b,d,\t$ are small
enough and that they satisfy a number of inequalities, some of which will now 
be discussed. Such inequalities will be satisfied by proper choices of 
$b$ and $\t$, as long as $d$ is small enough.

\begin{enumerate}
\item We require $\r R_0^3\gg 1$, that is $0<d<1/6$. Note that under this 
condition we also have $a_0/R_0 \gg \sqrt{\rho a_0^3} \gg (a_0/R_0)^3$. 
\item We require $\ell t R_0^{-1}\gg 1$, as in (\ref{2.6}), so that 
$b+d>\t$.
\item We require $a_0 R_0^{-1}\ell^{-2} t^{-1}\ll \r a_0\sqrt{\r a_0^3}$, 
as in (\ref{2.6}), so that $2b-d-\t>0$.
\item Noting that the contribution from the potential energy per particle in 
$H^n_\ell$ is expected (on the basis of Bogoliubov theory)
to be of order $n a_0 R_0^{-1}$ relative to the main term, 
we require both that $ta_0 R_0^{-1}$ and $a_0(R^{-1}-
R_0^{-1})$ are much smaller than $\sqrt{\r a_0^3}$, 
in order to guarantee that the errors produced
by the presence of $\g$ in front of the potential energy and by the replacement
of $R_0$ with $R$ are negligible. These conditions imply $\t>d$ and $2b+d>\t$.
\end{enumerate}

Further requirements will be discussed below.

\subsection{A priori bounds on $n$ and $n_+$}\label{ss:ap}

As a first step in our 
argument, we shall derive preliminary
bounds on the number of particles minimizing $E^n_\ell$ and on the 
average number of particles $\media{\hat n_+}$ outside the condensate. [Here
the operator $\hat n_+$ is defined, in second quantized form, as $\hat n_+=
\sum_{\pp\neq{\bf 0}}a^{\dagger}_\pp a_\pp$.] First of all, note that $0\ge 
\inf_{1\le n\le N} E^n_\ell$, so we can restrict our attention to 
the values of $n$ such that $E^n_\ell\le 0$. As proved in the next lemma, 
such values of $n$ cannot be too small, namely they all satisfy 
$n\ge c\r\ell^3$ for a suitable constant $c$. We can thus assume, without loss of generality, 
that $n\ge c\r\ell^3$ in the following.

{\lemma\label{lemma2} 
If $\ell/R$ and $t^{-1}$ are large enough, then $H^n_\ell\geq 0$ if 
$n\leq \rho\ell^3/4$.}\\

{\cs Proof.} From the definition of $H^n_\ell$ we see immediately that
\be H^n_\ell\geq\frac{\g a_0 R}{R_0^4}
\Big[-\r \sum_{j=1}^n \int\,d\yy w_R(\xx_i,\yy)+\frac{\r^2}{2}\int\!\!\!\int\,d\xx 
d\yy w_R(\xx,\yy)\Big]\ge 
2\pi\frac{\g a_0 R^4}{R_0^4}\big(-4 n\r +\r^2 \ell^3 \big)\;,
\lb{2.20}\ee
where we used that $\sup_{\xx}\int w_{R}(\xx,\yy)\,d\yy\leq 8\pi R^3$ 
and that, for $\ell/R$ and $t^{-1}$ large enough,
$\int\!\!\!\int\,d\xx 
d\yy w_R(\xx,\yy)\ge 4\pi R^3 \ell^3$. This proves the lemma.\qed\\

A similar argument allows us to get a preliminary bound on the average number 
of particles outside the condensate $\media{\hat n_+}$.

{\lemma\label{III.3} If $t$ and $(R_0-R)R_0^{-1}$ are small enough, 
then for any state such that 
$\media{H^n_\ell}\le 0$, the expectation of the number of excited particles 
satisfies $\media{\hat n_+}\le \const n a_0\ell^2 R_0^{-3}$.}\\

{\cs Proof.} Using the fact that the potential $e^{-|\xx|/R}$ is positive 
definite, we obtain
\be H^n_\ell\ge -\sum_{i=1}^n\D^{(j)}_\ell-\frac{\g a_0 R}{2 R_0^4}\sum_{i=1}^n
w_R(\xx_i,\xx_i)\;.
\lb{2.21}\ee
The claim of the lemma follows by using 
$\media{-\sum_{i=1}^n\D^{(j)}_\ell}\ge \media{\hat n_+}\p^2/\ell^2$
and the fact that $w_R(\xx,\xx)\le 1$.\qed\\

Of course, in order for the bound in Lemma \ref{III.3} to be useful, 
it must be $a_0\ell^2 R_0^{-3}\ll 1$. In the following we shall impose this 
condition by requiring that, in terms of the scaling exponents introduced 
in the previous section, $2b+3d<1/2$. We shall define $\n_0=1/2-2b-3d$,
so that our preliminary a priori bound reads $\media{\hat n_+}/n\le Y^{\n_0}$.

\subsection{Bound on the unimportant part of the Hamiltonian}
\label{unimportant}

Motivated by Bogoliubov's computation of the ground state energy, we would like
to be able to neglect in $H^n_{\ell}$ all terms but
those containing precisely two $a^\#_\pp$, with $\pp\neq{\bf 0}$. 
Let $\hat n_\V0=a^{\dagger}_\V0 a_\V0$ and let us rewrite (\ref{2.18}) in 
the form
\bea &&H^n_{\ell}=
\sum_{\pp} \pp^2a^{\dagger}_\pp a_\pp+\frac{\g a_0 R}{2R_0^4}
\sum_{\pp,\qq\neq\V0}\hw_{\pp\qq,\V0\V0}\left(a^{\dagger}_\pp 
a^{\dagger}_\qq a_\V0 a_\V0
+2a^{\dagger}_\pp a^{\dagger}_\V0 a_\V0 a_\qq+
a^{\dagger}_\V0 a^{\dagger}_\V0 a_\pp a_\qq\right)+
\nonumber\\
&+&
\frac{\g a_0 R}{2R_0^4}\Bigl[
\hw_{\V0\V0,\V0\V0}\left[(\hat n_0-\r\ell^3)^2-\hat n_0\right]+
2\sum_{\pp\neq\V0}\hw_{\pp\V0,\V0\V0}
\left[(\hat n_0-\r\ell^3)a^{\dagger}_\pp a_\V0
+a^{\dagger}_\V0 a_\pp(\hat n_0-\r\ell^3)\right]+\nonumber\\
&+&
2\sum_{\pp,\qq\neq\V0}\hw_{\pp\V0,\qq\V0}a^{\dagger}_\pp a_\qq
(\hat n_0-\r\ell^3)
+2\sum_{\pp,\qq,\mm\neq\V0}\hw_{\pp\qq,\mm\V0}\left(
a^{\dagger}_\pp a^{\dagger}_\qq a_\mm a_\V0+
a^{\dagger}_\V0 a^{\dagger}_\mm a_\qq a_\pp\right)+\nonumber\\
&+&\sum_{\pp,\qq,\mm,\Bn\neq\V0}\hw_{\pp\qq,\mm\Bn}a^{\dagger}_\pp 
a^{\dagger}_\qq a_\mm a_\Bn\Bigr]\;.
\lb{2.22}\eea
We would like to show that all terms but those in the first line are 
negligible. Let us then estimate all these contributions in terms of
$\hat n_+$ and $n-\r\ell^3$. We shall use, without proof, a number of lemmas
from \cite{LS}. Note that from now on we shall always assume valid the conditions discussed
in Sec.~\ref{scalings} above. Note also that  $\hat n_0=n-\hat n_+$ and 
$|\hw_{\pp\qq,\mm\Bn}|
\le \const R^3/\ell^3$. 

\begin{enumerate}
\item The first term in the second line of (\ref{2.22}) satisfies 
\be\frac{\g a_0 R}{2R_0^4}
\hw_{\V0\V0,\V0\V0}\left[(\hat n_0-\r\ell^3)^2-\hat n_0\right]\ge
\const\frac{a_0}{\ell^3}(\hat n_0-\r\ell^3)^2-\const \frac{na_0}{\ell^3}
\lb{2.23}\ee
and, for any $\e>0$
\be\frac{\g a_0 R}{2R_0^4}
\hw_{\V0\V0,\V0\V0}\left[(\hat n_0-\r\ell^3)^2-\hat n_0\right]\ge
\frac{\g a_0 R}{2R_0^4}
\hw_{\V0\V0,\V0\V0}(1-\e)(n-\r\ell^3)^2-\const \frac{a_0}{\ell^3}
\frac{\hat n_+^2}\e-\const \frac{a_0n}{\ell^3}\lb{2.23a}\ee
\item By Lemma 5.5 of \cite{LS}, the second term in the second line of 
(\ref{2.22}) 
satisfies for any $\e>0$
\bea
&&\frac{\g a_0 R}{R_0^4}
\sum_{\pp\neq\V0}\hw_{\pp\V0,\V0\V0}
\left[(\hat n_0-\r\ell^3)a^{\dagger}_\pp a_\V0
+a^{\dagger}_\V0 a_\pp(\hat n_0-\r\ell^3)\right]\lb{2.24}\\
&&\qquad\ge-\const \frac{\hat n_+}{\e}
\frac{na_0}{\ell^3}-\const \e\frac{a_0}{\ell^3}(\hat n_0-\r\ell^3)^2-\const
\e \frac{a_0}{\ell^3}\nonumber  \eea
and
\bea
&&\frac{\g a_0 R}{R_0^4}
\sum_{\pp\neq\V0}\hw_{\pp\V0,\V0\V0}\left[(\hat n_0-\r\ell^3)
a^{\dagger}_\pp a_\V0
+a^{\dagger}_\V0 a_\pp(\hat n_0-\r\ell^3)\right]\lb{2.24a}\\
&&\quad\ge\frac{\g a_0 R}{R_0^4}
\sum_{\pp\neq\V0}\hw_{\pp\V0,\V0\V0}\left[(n-\r\ell^3)a^{\dagger}_\pp a_\V0
+a^{\dagger}_\V0 a_\pp(n-\r\ell^3)\right]-\const \e\hat n_+ 
\frac{a_0 n}{\ell^3}
-\const \frac{\hat n_+^2}{\e}\frac{a_0}{\ell^3}\nonumber\eea
\item By Lemma 5.3 of \cite{LS} the first term in the third line of 
(\ref{2.22}) 
satisfies
\be\frac{\g a_0 R}{R_0^4}
\sum_{\pp,\qq\neq\V0}\hw_{\pp\V0,\qq\V0}a^{\dagger}_\pp 
a_\qq(\hat n_0-\r\ell^3)
\ge -\const \frac{a_0}{\ell^3}[\r\ell^3-n]_+\hat n_+-\const \frac{a_0
\hat n_+^2}{\ell^3}\lb{2.25}\ee
where $[t]_+=\max\{t,0\}$.
\item By Lemma 5.6 of \cite{LS} and its proof, the second term in the 
third line of 
(\ref{2.22}) satisfies for any $\e>0$
\bea&&\frac{\g a_0 R}{R_0^4}
\sum_{\pp,\qq,\mm\neq\V0}\hw_{\pp\qq,\mm\V0}\left(
a^{\dagger}_\pp a^{\dagger}_\qq a_\mm a_\V0+a^{\dagger}_\V0 a^{\dagger}_\mm 
a_\qq a_\pp\right)\lb{2.26}\\
&&\qquad\ge
-\const \e \frac{a_0 n}{\ell^3}\hat n_+-\const\frac1\e
\frac{a_0\hat n_+}{R^3}
-\frac1\e\frac{\g a_0 R}{R_0^4}\sum_{\pp,\qq,\mm,\Bn\neq\V0}
\hw_{\pp\qq,\mm\Bn}a^{\dagger}_\pp a^{\dagger}_\qq a_\mm a_\Bn\nonumber\eea
\item The term in the fourth line of (\ref{2.22}) satisfies
\be 0\le\frac{\g a_0 R}{2R_0^4}\sum_{\pp,\qq,\mm,\Bn\neq\V0}
\hw_{\pp\qq,\mm\Bn}a^{\dagger}_\pp a^{\dagger}_\qq 
a_\mm a_\Bn\le \const\frac{a_0\hat n_+^2}{R^3}\;.\lb{2.27}\ee
This follows immediately from the fact that $w_{R}(\xx,\yy)\le 1$.
\end{enumerate}
\\

{\bf Remark.} According to Bogoliubov's theory we expect that in the ground state
$\media{\hat n_+}\sim n\sqrt{\r a_0^3}$. From  
the upper bound in (\ref{2.27}) we thus expect that the contribution
to the ground state energy from the quartic term
$\frac{\g a_0 R}{2R_0^4}\sum_{\pp,\qq,\mm,\Bn\neq\V0}
\hw_{\pp\qq,\mm\Bn}a^{\dagger}_\pp a^{\dagger}_\qq a_\mm a_\Bn$ is at most 
$\sim n^2\r a_0\frac{
a_0^3}{R^3}$. In order to show that Bogoliubov theory is asymptotically
correct up to terms of order $n\r a_0\sqrt{\r a_0^3}$ we shall require
such a bound to be much smaller than $n\r a_0\sqrt{\r a_0^3}$. For 
$n\sim\r\ell^3$ and in terms of the scaling exponents introduced above, 
this implies $Y^{1/2-3b-3d}\ll 1$, that is $3b+3d<1/2$. In the following
we shall assume this condition valid. It will be convenient
to summarize here all the requirement we asked for so far on the scaling
exponents introduced in Sec.~\ref{scalings}:
\be 2b-d>\t>d\;,\qquad \text{and} \quad  \frac16>b+d>\t\;.\lb{2.27b}\ee
{}From now on we shall always assume that
these relations are valid and that $Y$ is small enough.

\subsection{The quadratic Hamiltonian}\label{quadratic}

In this section we consider the main part of the Hamiltonian. 
This is the ``quadratic'' Hamiltonian considered by Bogoliubov.
It consists of the kinetic energy and all the terms with the 
coefficients $\hw_{\pp\qq,\V0\V0}$, $\hw_{\V0\V0,\pp\qq}$ $\hw_{\pp\V0,
\V0\qq}$, and $\hw_{\V0\pp,\qq\V0}$ with $\pp,\qq\ne\V0$, i.e.,
\be H_B=\sum_{\pp} \pp^2a^{\dagger}_\pp a_\pp+\frac{\g a_0 R}{2R_0^4}
\sum_{\pp,\qq\neq\V0}\hw_{\pp\qq,\V0\V0}\left(a^{\dagger}_\pp 
a^{\dagger}_\qq a_\V0 a_\V0
+2a^{\dagger}_\pp a^{\dagger}_\V0 a_\V0 a_\qq+a^{\dagger}_\V0 a^{\dagger}_\V0 
a_\pp a_\qq\right)\;.\lb{2.28}\ee
In order to compute all  the bounds we find it necessary to include 
the first term in the second line of (\ref{2.24a}) into the ``quadratic''
Hamiltonian. We therefore define 
\bea H_{Q}&=&
\sum_{\pp} \pp^2a^{\dagger}_\pp a_\pp+\frac{\g a_0 R}{R_0^4}
\sum_{\pp\neq\V0}\hw_{\pp\V0,\V0\V0}\left[(n-\r\ell^3)a^{\dagger}_\pp a_\V0
+a^{\dagger}_\V0 a_\pp(n-\r\ell^3)\right]\lb{2.29} \\
&+&\frac{\g a_0 R}{2R_0^4}
\sum_{\pp,\qq\neq\V0}\hw_{\pp\qq,\V0\V0}\left(a^{\dagger}_\pp 
a^{\dagger}_\qq a_\V0 a_\V0
+2a^{\dagger}_\pp a^{\dagger}_\V0 a_\V0 a_\qq+a^{\dagger}_\V0 
a^{\dagger}_\V0 a_\pp a_\qq\right)\;.\nonumber\eea
Note that $H_{B}=H_Q$ in the neutral case $n=\rho\ell^3$.
Our goal is to give a lower bound on the ground state energy of the
Hamiltonian $H_Q$.

For any $\kk\in\RRR^3$ denote $\chi_{\ell,\kk}(\xx)=e^{i\kk\xx}\chi(\xx/\ell)$ 
and define the operators
\be b_\kk^{\dagger}=\sum_{\pp\neq\V0}(\phi_\pp,\c_{\ell,\kk})a^{\dagger}_\pp
a_\V0\quad
\hbox{and}\quad
b_\kk=\sum_{\pp\neq\V0}(\c_{\ell,\kk},\phi_\pp)a^{\dagger}_\V0 a_\pp\;.
\lb{2.30}\ee
Note that they satisfy the commutation relations
\be[b_\kk,b^\dagger_{\kk'}]=\hat n_\V0(\c_{\ell,\kk},\c_{\ell,\kk'})
-\sum_{\pp,\qq\neq\V0}(\c_{\ell,\kk},\phi_\pp)(\phi_\qq,\c_{\ell,\kk'})
a^{\dagger}_\qq
a_\pp-\hat n_\V0(\c_{\ell,\kk},\phi_\V0)(\phi_\V0,\c_{\ell,\kk'})
\;.\lb{2.31}\ee
Using Lemma 6.2 of \cite{LS} we find that 
\be\left\langle\sum_\pp|\pp|^2 a^{\dagger}_\pp a_\pp\right\rangle\geq 
(1-C't)^2n^{-1}\int_{\rrr^3}\,\frac{d\kk}{(2\p)^3}\, 
\frac{|\kk|^4}{|\kk|^2+(\ell t^3)^{-2}}\,
\langle b_\kk^{\dagger}b_\kk\rangle\lb{2.32}\ee
for a suitable constant $C'$ and for all states with 
particle number equal to $n$.
Concerning the potential energy terms, note that we may write
\be w_{R}(\xx,\yy)=\int_{\rrr^3}\,\frac{d\kk}{(2\p)^3}\,\hat{V}_{R}(\kk)
\chi_{\ell,\kk}(\xx)\chi^*_{\ell,\kk}(\yy)\;,\lb{2.33}\ee
where $\hat{V}_{R}(\kk)=8\p R^3[1+(\kk R)^2]^{-2}$.
The last two sums in the Hamiltonian (\ref{2.29}) can  therefore be written as
\bea&&\frac{\g a_0 R}{R_0^4\ell^3}\int_{\rrr^3}\,\frac{d\kk}{(2\p)^3}
\hat{V}_{R}(\kk)\Bigl[(n-\rho\ell^3)
\ell^{3/2}\left(\hat{\chi}(\kk \ell)b^{\dagger}_\kk+
\hat{\chi}^*(\kk\ell)b_\kk\right)\lb{2.34}\\ 
&&+\frac{1}{2}\left(b^{\dagger}_\kk b_\kk+b^{\dagger}_{-\kk}b_{-\kk}+
b^{\dagger}_{\kk}b^{\dagger}_{-\kk}
+b_\kk b_{-\kk}\right)\Bigr]\,-\frac{\g a_0 R}
{R_0^4}\sum_{\pp,\qq\neq\V0}\hw_{\pp\qq,\V0\V0}a^{\dagger}_\pp a_\qq\;.
\nonumber\eea
Thus, we have for states with particle number equal to $n$ that 
\be\left\langle{H}_Q\right\rangle 
\geq \int_{\rrr^3}\,\frac{d\kk}{(2\p)^3}\,\left\langle h_Q(\kk)\right\rangle
-\frac{\g a_0 R}{R_0^4}\sum_{\pp,\qq\neq\V0}
\hw_{\pp\qq,\V0\V0}\left\langle a^{\dagger}_\pp a_\qq\right\rangle,\lb{2.35}\ee
where
\bea h_Q(\kk)&=&\frac{(1-C't)^2}{2 n}
\frac{|\kk|^4}{|\kk|^2+(\ell t^3)^{-2}}
\left(b_\kk^{\dagger}b_\kk+b_{-\kk}^{\dagger}b_{-\kk}\right)\lb{2.36}\nonumber\\ 
&&+\frac{\g a_0 R}{2R_0^4\ell^3}\,\hat{V}_{R}(\kk)\,\Bigl[(n-\rho\ell^3)
\ell^{3/2}\left(\hat{\chi}(\kk\ell)(b^{\dagger}_\kk+b_{-\kk})+
\hat{\chi}^*(\kk\ell)(b_\kk+b^{\dagger}_{-\kk})\right)\\
&&\qquad\qquad\qquad\quad +b^{\dagger}_\kk b_\kk+b^{\dagger}_{-\kk}b_{-\kk}+b^{\dagger}_{\kk}
b^{\dagger}_{-\kk}+b_\kk b_{-\kk}\Bigr]\;.\nonumber\eea
In order to give a lower bound on $h_Q(\kk)$, we can use 
Bogoliubov's 
method, in form of Theorem 6.3 of \cite{LS}. This theorem states that, 
for arbitrary constants $\cA\ge\cB>0$ and $\k\in\CCC$, the inequality
\bea&&\cA(b^{\dagger}_\kk b_\kk+b^{\dagger}_{-\kk}b_{-\kk})+\cB(b^{\dagger}_\kk
b^{\dagger}_{-\kk}+b_\kk b_{-\kk})+
\kappa(b^{\dagger}_\kk+b_{-\kk})+\kappa^*(b_\kk+b^{\dagger}_{-\kk})\lb{2.37}\\ 
&&\geq-\frac{1}{2}\big(\cA-\sqrt{\cA^2-\cB^2}\big)
([b_{\kk},b^{\dagger}_{\kk}]+[b_{-\kk},b^{\dagger}_{-\kk}])-
\frac{2|\kappa|^2}{\cA+\cB}\nonumber\eea
holds. Note that in our case
\be[b_\kk,b_{\kk'}^\dagger]\le \hat n_\V0
\int\,d\xx\,\c(\xx/\ell)^2\le n\ell^3\;.\lb{2.38}\ee
With the notation 
\bea&&\cB_\kk=\frac{\g a_0 R}{2R_0^4\ell^3}\hat{V}_{R}(\kk)\;,\nonumber\\
&&\cA_\kk=\frac{(1-C't)^2}{2n}\,\frac{|\kk|^4}{|\kk|^2+(\ell t^3)^{-2}}
+\cB_\kk\;,\lb{2.40}\\
&&\k_\kk=\frac{\g a_0 R}{2R_0^4\ell^{3/2}}\hat{V}_{R}(\kk)
(n-\r\ell^3)
\hat\c(\kk\ell)\nonumber\eea
we thus obtain that on the subspace of $n$ particles
\be h_Q(\kk) \geq - n \ell^3 \left(\cA_\kk-\sqrt{\cA_\kk^2-\cB_\kk^2}\right)
- \frac{2|\kappa_\kk|^2}{\cA_\kk+\cB_\kk} \;. \label{lql} \ee
Moreover, since  
\bea 
\sum_{\pp,\qq\neq\V0}\hw_{\pp\qq,\V0\V0}a^{\dagger}_\pp a_\qq&=&
\int \frac{d\xx}{\ell^3}\int d\yy
w_R(\xx,\yy)\Big[\sum_{\pp\neq\V0}\phi_\pp(\xx)a_\pp\Big]^\dagger
\Big[\sum_{\pp\neq\V0}\phi_\pp(\yy)a_\pp\Big]\le\label{2.38a}\\
&\le&\int \frac{d\xx}{\ell^3}\int d\yy
w_R(\xx,\yy)\Big[\sum_{\pp\neq\V0}\phi_\pp(\xx)a_\pp\Big]^\dagger
\Big[\sum_{\pp\neq\V0}\phi_\pp(\xx)a_\pp\Big]\le \frac{8\p R^3}{\ell^3}\hat n_+
\nonumber\eea
we have that 
\be \frac{\g a_0R}{R_0^4}\sum_{\pp,\qq\neq\V0}
\hw_{\pp\qq,\V0\V0}a^{\dagger}_\pp a_\qq\le \const n\frac{a_0}{\ell^3}\;.
\label{2.38b}\ee
Using (\ref{2.35}), (\ref{lql}) and (\ref{2.38b}), we find that, 
on the subspace with $n$ particles, 
\be H_Q\ge -\int_{\rrr^3}\,\frac{d\kk}{(2\p)^3}\,
\Bigl\{n\ell^3\left(\cA_\kk-\sqrt{\cA_\kk^2-\cB_\kk^2}\right)
+\frac{2|\k_\kk|^2}{\cA_\kk+\cB_\kk}\Bigr\}-\const n\frac{a_0}{\ell^3} \;. \lb{2.39}
\ee

Now, using $\cA_\kk\ge\cB_\kk$ and the definitions of $\cB_\kk,\k_\kk$, we get 
\bea&&\int_{\rrr^3}\,\frac{d\kk}{(2\p)^3}
\frac{2|\k_\kk|^2}{\cA_\kk+\cB_\kk}\le \int_{\rrr^3}\,\frac{d\kk}{(2\p)^3}
\frac{|\k_\kk|^2}{\cB_\kk}\lb{2.41}\\
&&=\frac{\g a_0 R}{2R_0^4}(n-\r\ell^3)^2
\int_{\rrr^3}\,\frac{d\kk}{(2\p)^3}\hat{V}_{R}(\kk)
|\hat\c(\kk\ell)|^2
=\frac{\g a_0 R}{2R_0^4}(n-\r\ell^3)^2\hat
w_{\V0\V0,\V0\V0}\;.\nonumber\eea
As a result, on the subspace with $n$ particles,
\be H_Q\ge -n I-\frac{\g a_0 R}{2R_0^4}(n-\r\ell^3)^2\hat
w_{\V0\V0,\V0\V0}-\const\,n\frac{a_0}{\ell^3}\;,\lb{2.42}\ee
where 
\bea&&I=\frac1{2\r}\int\,\frac{d\kk}{(2\p)^3}\,\left[
f(\kk)-\sqrt{f(\kk)^2-g(\kk)^2}\right]\;,\lb{2.43}\\
&& f(\kk)=(1-C't)^2\frac{\r\ell^3}{n}\,\frac{|\kk|^4}{|\kk|^2+(\ell t^3)^{-2}}
+\frac{\g a_0 R\r}{R_0^4}\hat V_{R}(\kk)\;,\nonumber\\
&&g(\kk)=\frac{\g a_0 R\r}{R_0^4}\hat V_{R}(\kk)\;.\nonumber\eea
Similarly, the Bogoliubov Hamiltonian in (\ref{2.28}) on 
the subspace with $n$ particles admits the lower bound
\be H_B\ge -n I-\const\,n\frac{a_0}{\ell^3}\;.\lb{2.44}\ee

Note that $f> g> 0$ implies $f-\sqrt{f^2-g^2}\le \min\{g,
g^2/(f-g)\}$. Thus clearly $I$ can be bounded as $I\le (2\p)^{-3}(2\r)^{-1}
\int d\kk \, g(\kk)\le \const a_0 R^{-3}$, so that 
\be H_B\ge -\,\const n\frac{a_0}{R^3}\;.\lb{2.46}\ee
Moreover, if $n\le C\r\ell^3$, we find
\bea I&&\le\frac1{2\r}\left\{\int_{|\kk|^2\le a_0\r}\,
\frac{d\kk}{(2\p)^3}\,g(\kk)+ \int_{|\kk|^2\ge a_0\r}\,
\frac{d\kk}{(2\p)^3}\,\frac{g(\kk)^2}{f(\kk)-g(\kk)}\right\}\nonumber\\
&&\le \const\left\{\r a_0 \sqrt{\r a_0^3}+\r a_0^2\int_{
\sqrt{\r a_0}}^\io\,dk\,\frac1{[(kR)^2+1]^4}\,\left(
1+(k\ell t^3)^{-2}\right)\right\}\lb{2.45} \nonumber \\
&&\le\const \r a_0\left\{\sqrt{\r a_0^3}+\frac{a_0}{R}\left[
1+\frac1{\sqrt{\r a_0}R}\left(\frac{R}{\ell t^3}\right)^2\right]\right\}\;.
\eea
If the scaling exponents satisfy 
\be 2b+d-6\t>0\;,\label{2.46ab}\ee
then the last expression in (\ref{2.45}) can be bounded from above by 
$\const \r a_0\frac{a_0}R$. Hence, if $n\le C\r\ell^2$ and $2b+d-6\t>0$, 
\be H_B\ge -\,\const n\r a_0\frac{a_0}{R}\;.\lb{2.46aa}\ee

\subsection{Improved bounds on $n$}

Using the bounds derived in the previous sections, 
we shall now get an improved 
bound on $n$, which implies that for states with $\media{H^n_\ell}\le 0$, $n$ cannot 
deviate too much from $\r\ell^3$. 
In order to bound (\ref{2.22})
from below, we use (\ref{2.23}), (\ref{2.24}), (\ref{2.25}), (\ref{2.26}) and
(\ref{2.46}) [note that we shall use (\ref{2.26}) with $\e$ replaced 
by $\e^{-1}$]. The result is that, for some positive 
constants $c$ and $C$, we have
\bea&&H^n_\ell\ge (c-C \e)\frac{a_0}{\ell^3}(\hat n_0-\r\ell^3)^2
+(1-C\e)\frac{\g a_0 R}{2R_0^4}\sum_{\pp,\qq,\mm,\Bn\neq\V0}
\hw_{\pp\qq,\mm\Bn}a^{\dagger}_\pp a^{\dagger}_\qq a_\mm a_\Bn\lb{2.47}\\
&&-C\left\{\frac{na_0}{\ell^3}
+\frac{\hat n_+}{\e}\frac{na_0}{\ell^3}
+\e \frac{a_0}{\ell^3}
+\frac{a_0}{\ell^3}\r\ell^3\hat n_+
+\frac{a_0\hat n_+^2}{\ell^3}
+\e\frac{a_0\hat n_+}{R^3}+n\frac{a_0}{R^3}\right\}\nonumber\eea
for some $\e>0$. 
Choosing $\e=\min\{1,c\}/(2C)$, using $\hat n_+\le n$, 
$\sum_{\pp,\qq,\mm,\Bn\neq\V0}
\hw_{\pp\qq,\mm\Bn}a^{\dagger}_\pp a^{\dagger}_\qq a_\mm a_\Bn\ge 0$ 
and recalling
that, by Lemma \ref{lemma2}, $n\le \r\ell^3/4$ implies $H^n_\ell\ge 0$,
we have, for some new constants $c'$ and $C'$ and for any state with 
$\media{H^n_\ell}\le 0$, 
\be 0\ge\media{H^n_\ell}
\ge c'\frac{a_0}{\ell^3}\media{(\hat n_0-\r\ell^3)^2}
-C'\frac{a_0}{\ell^3}\left\{n\media{\hat n_+}+\frac{n^2}{\r R^3}
\right\}\lb{2.48}\ee
and, therefore,
\be \frac{(n-\r\ell^3)^2}{n^2}\le \const
\left\{\frac{\media{\hat n_+}}{n}
+\frac{1}{\r R^3}\right\}\;.\lb{2.49}\ee
Here, we used $\media{(\hat n_0-\r\ell^3)^2}\ge (n-\r\ell^3)^2-2n\media{
\hat n_+}$. Now, let us recall from Section~\ref{ss:ap} that, in terms of the scaling exponents $b,d$,
we have $\media{\hat n_+}/n
\le Y^{\frac12-2b-3d}$, 
and $(\r R^3)^{-1}\sim
Y^{\frac12-3d}$, so that
\be\frac{(n-\r\ell^3)^2}{n^2}\le \const Y^{\n_0}\;,\lb{2.50}\ee
where $\n_0=1/2-2b-3d$ as before. Eq.~(\ref{2.50})
can be rewritten as 
\be \left|n-\r\ell^3\right|\le \const\,\r\ell^3\,Y^{\n_0/2}\lb{2.50a}\;.\ee

In order to get the bounds above we sacrificed all the kinetic
energy in (\ref{2.22}). Of course this is not necessary: we can decide to 
sacrifice only half of it and we would still get the same 
bounds, only with different constants. If we proceed in this way we see 
that for any $n$-particle state such that 
$\media{H^n_\ell}\le 0$,
\be \sum_\pp|\pp|^2\media{a^{\dagger}_\pp a_\pp}\le\const n\r a_0
Y^{\n_0} \;.\lb{2.51}\ee

\subsection{Localization of $n_+$}

The idea now is to use the improved bound on $n$ together with the
bounds in previous sections in order to find an improved bound on the
energy of the ground state. In order to do this it is clear from the
bounds in Sec.~\ref{unimportant} that we need to estimate $\media{\hat
  n_+^2}$.  Since we have bounded only $\hat n_+$ so far, we would
like to argue that $\langle \hat n_+^2 \rangle \approx \langle \hat
n_+\rangle^2$. In this section we shall discuss how to do this. We
shall utilize the following theorem, which is Theorem~A.1 of
\cite{LS}. [The $k^{\rm th}$ supra- (resp. infra-) diagonal of a
matrix ${\cal A}$ is the submatrix consisting of all elements $a_{i,
  i+k}$ (resp. $a_{i+k , i}$)].

{\theorem\label{thmIII.1} 
Suppose that ${\cal A}$ is an $N\times N$ Hermitean matrix and let
${\cal A}^k$, with $k=0,1,\ldots,N-1$, denote the matrix consisting of
the $k^{\rm th}$ supra- and infra-diagonal of ${\cal A}$.  Let $\psi
\in {\bf C}^N$ be a normalized vector and set $d_k = (\psi , {\cal
A}^k \psi) $ and $\lambda = (\psi , {\cal A} \psi) =
\sum_{k=0}^{N-1} d_k$  \ ($\psi$ need not be an eigenvector of
${\cal A}$). Choose some positive integer $M \leq N$.  Then, with $M$ fixed,
there is some $n \in [0, N-M]$ and some normalized vector $ \phi \in
{\bf C}^N$ with the property that $\phi_j =0$ unless $n+1 \leq j
\leq n+M$ \ (i.e., $\phi $ has length $M$) and such that
\be(\phi , {\cal A} \phi) \leq \lambda+\frac{C}{ M^2}
\sum_{k=1}^{M-1} k^2 |d_k|+C\sum_{k=M}^{N-1}|d_k|\lb{2.52}\ee
where $C>0 $ is a  universal constant. (Note that the first sum starts with 
$k=1$).}\\

{}From this theorem we can get a localization bound on $\hat n_+$ in the
following way.  Consider a normalized $n$-particle wavefunction
$\Psi$, which we may write as $\Psi=\sum_{m=0}^n c_m\Psi_m$, where for
all $m=0,1,2,\ldots,n$, $\Psi_m$ is a normalized eigenfunction of
$\hat n_+$ with eigenvalue $m$.  We now consider the
$(n+1)\times(n+1)$ Hermitean matrix ${\cal A}$ with matrix elements
${\cal A}_{mm'}=\left(\Psi_m,H^n_{\ell} \Psi_{m'}\right)$.

We shall use Theorem \ref{thmIII.1} for this matrix and the vector 
$\psi=(c_0,\ldots,c_n)$. We shall choose $M$ in  Theorem \ref{thmIII.1} 
to be of the order of the upper bound on $\langle\hat n_+\rangle$ derived in 
Lemma \ref{III.3}, e.g., $M$ is the integer part of $n Y^{\n_0}$.
Note that, if $n\sim\r\ell^3$, we have $M\gg1$.  
With the notation in Theorem \ref{thmIII.1} we have 
$\lambda=(\psi,{\cal A}\psi)=
(\Psi,H^n_{\ell}\Psi)$. 
Note also that because of the structure of $H^n_{\ell}$ 
we have, again with  the notation from Theorem \ref{thmIII.1}, 
that $d_k=0$ if $k\ge 3$.
We conclude from it that there exists a normalized 
wavefunction $\widetilde{\Psi}$ with the property that the corresponding
$\hat n_+$ values belong to an interval of length $M\sim nY^{\n_0}$ 
and such that 
\be\left(\Psi,H^n_{\ell}\Psi\right)
\geq \left(\widetilde{\Psi},H^n_{\ell}\widetilde{\Psi}\right)
-\const \frac1{n^2 Y^{2\n_0}}(|d_1|+|d_2|).\lb{2.53}\ee

We shall now bound $d_1$ and $d_2$. We have
$d_1=(\Psi,H^n_{\ell}(1)\Psi)$, where $H^n_{\ell}(1)$ is the part of
the Hamiltonian $H^n_{\ell}$ containing all the terms with the
coefficients $\hw_{\pp\qq,\mm\Bn}$ for which precisely one or three
indices are $\V0$.  These are the terms bounded in (\ref{2.24}) and
(\ref{2.26}).  These estimates are stated as one-sided bounds.  It is
however clear that they could have been stated as two sided
bounds. Using in addition the bound (\ref{2.27}) and $\hat n_+^2 \leq
n \hat n_+$ then, for any $\e>0$ and some positive constant $C$, we
get
\be |d_1| \leq C\Big(\Psi, \Big[
\frac{\hat n_+}{\e}\frac{na_0}{\ell^3}
+\e\frac{a_0}{\ell^3}(\hat n_0-\r\ell^3)^2
+\e \frac{a_0}{\ell^3}
+\e\frac{a_0n\hat n_+}{R^3}\Big]\Psi\Big) \,. \ee
If $\Psi$ satisfies $(\Psi, H^n_\ell\Psi)\le 0$, we can use 
Lemma~\ref{III.3} and (\ref{2.50a}) to conclude that 
\be |d_1|\le C\,n\r a_0\left(\frac{Y^{\n_0}}{\e}+\e Y^{\n_0-3b-
3d}\right)\;.\lb{2.55}\ee
Optimizing over $\e>0$ yields the bound 
\be |d_1|\le C n\r a_0\,Y^{\n_0-\frac32(b+d)}  =C n\r a_0\,Y^{\frac12-\frac72b-\frac92d}\;.\lb{2.56}\ee

For $d_2$ we obtain
\bea |d_2|&&\leq \Big(\Psi,\frac{\g a_0 R}{2R_0^4}
\sum_{\pp,\qq\neq\V0}\hw_{\pp\qq,\V0\V0}\left(
a^{\dagger}_\pp a^{\dagger}_\qq a_\V0 a_\V0
+a^{\dagger}_\V0 a^{\dagger}_\V0 a_\pp a_\qq\right)\Psi\Big)\nonumber\\
&&=\Big(\Psi,\Big[ \sum_{\pp} \pp^2a^{\dagger}_\pp a_\pp+
\frac{\g a_0 R}{R_0^4}
\sum_{\pp,\qq\neq\V0}\hw_{\pp\qq,\V0\V0}a^{\dagger}_\pp a^{\dagger}_\V0 
a_\V0 a_\qq\Big]
\Psi\Big)-\Big(\Psi,\widetilde H_B\Psi\Big)\lb{2.57}\eea
where 
\be \widetilde H_B=\sum_{\pp}\pp^2a^{\dagger}_\pp a_\pp+\frac{\g a_0 R}{2R_0^4}
\sum_{\pp,\qq\neq\V0}\hw_{\pp\qq,\V0\V0}\left(-a^{\dagger}_\pp a^{\dagger}_\qq 
a_\V0 a_\V0+2a^{\dagger}_\pp a^{\dagger}_\V0 a_\V0 a_\qq-a^{\dagger}_\V0 
a^{\dagger}_\V0 a_\pp a_\qq\right)\lb{2.58}\ee
is an operator unitarily equivalent to $H_B$. (It is obtained from it 
by replacing $a^{\dagger}_\pp,a_\pp$ by $-ia^{\dagger}_\pp,ia_\pp$, 
respectively.) Of course $\widetilde H_B$ satisfies the same lower bound (\ref{2.46aa}) as $H_B$.
It is not difficult to see that 
\be 
0\leq \frac{\g a_0 R}{R_0^4}\sum_{\pp,\qq\ne0}\hw_{\pp\qq,\V0\V0}\media{
a^{\dagger}_\pp a^{\dagger}_\V0 a_\V0 a_{\qq}}
\leq\frac{4\p\g a_0}{\ell^3}\frac{R^4}{R_0^4}n\media{\hat n_+}\lb{2.58a}\ee
(compare with Lemma 5.4 of \cite{LS}). 
If $\Psi$ satisfies $(\Psi, H^n_\ell\Psi)\le 0$ 
then, using (\ref{2.57}), (\ref{2.58a}), 
Lemma \ref{III.3}, (\ref{2.51}) and (\ref{2.46aa}), we get:
\be |d_2|\le \const n\r a_0\left\{Y^{\n_0}+\frac{a_0}{R}\right\}\le 
\const n\r a_0\,Y^{\n_0}\;. \lb{2.59}\ee

Putting together these bounds we find that if $(\Psi, H^n_\ell\Psi)\le 0$
then there exists a normalized 
wavefunction $\widetilde{\Psi}$ with the property that the corresponding
$\hat n_+$ values belong to an interval of length $M\sim nY^{\n_0}$ 
and such that 
\bea\left(\Psi,H^n_{\ell}\Psi\right)&&\geq \left(\widetilde{\Psi},
H^n_{\ell}\widetilde{\Psi}\right)-\const\frac{\r a_0}{n Y^{2\n_0}}
Y^{\n_0-\frac32(b+d)}
\nonumber\\
&&\ge \left(\widetilde{\Psi},
H^n_{\ell}\widetilde{\Psi}\right)-Cn\r a_0\,
Y^{\m_0}\;,\lb{2.60}\eea
where $\m_0=-\n_0+1+\frac92b-\frac32d=
\frac{1}{2}+\frac{13}2b+\frac32d$. Since $\mu_0>1/2$, the error term
in the last line is much smaller than $n\r a_0\sqrt{\r
  a_0^3}$. Without loss of generality, we may assume that
$\left(\widetilde{\Psi},H^n_{\ell}\widetilde{\Psi}\right)\leq 0$, in
which case Lemma \ref{III.3} implies that $\left(\widetilde{\Psi},\hat
  n_+\widetilde{\Psi}\right)\le \const n Y^{\n_0}$. We also know that
the possible $\hat n_+$ values of $\widetilde{\Psi}$ range in an
interval of length $M\sim n Y^{\n_0}$. This implies that if
$\left(\Psi,H^n_{\ell}\Psi\right)\le -Cn\r a_0 Y^{1/2}$ then the
allowed values of $\hat n_+$ for $\widetilde{\Psi}$ are less than $C n
Y^{\n_0}$, for a suitable constant $C$. In particular, $\langle \hat
n_+^2\rangle \leq C n^2 Y^{2\nu_0}$ in the state $\widetilde
\Psi$. Hence, as far as the derivation of a lower bound on the ground
state energy is concerned, it is not a restriction to assume that
$\media{\hat n_+^2} \sim\media{\hat n_+}^2$. This fact will be used
in the next section to derive improved lower bounds on the ground
state energy.

\subsection{Improved bound on the ground state energy}

Let $\Psi$ be the ground state of $H^n_\ell$.
In this section we shall get an improved lower bound on $(\Psi,H^n_\ell\Psi)$
under the assumption that $(\Psi,H^n_\ell\Psi)$ is small enough such that (\ref{2.60}) implies that $(\widetilde \Psi, H_\ell^n \widetilde \Psi)\leq 0$. Note that
if this assumption is violated then the desired bound on the ground state 
energy would automatically be true. Hence, as discussed
at the end of previous section we know that 
\be(\Psi,H^n_\ell\Psi)\ge \media{H^n_{\ell}}-
C n \r a_0\,Y^{\m_0}\lb{2.60a}\ee
where the average 
$\media{\cdot}$ is over an $n$-particle state with allowed values of $\hat n_+$
smaller than $CnY^{\n_0}$. Using (\ref{2.22}),
(\ref{2.23a}), (\ref{2.24a}), (\ref{2.25}), (\ref{2.26}) [this time precisely
in the form stated, without replacement of $\e$ by $\e^{-1}$] and (\ref{2.27}),
we find that
\bea&&\media{H^n_{\ell}}\ge\media{H_Q}
+\frac{\g a_0 R}{2R_0^4}\hw_{\V0\V0,\V0\V0}(1-\e)(n-\r\ell^3)^2-C\Bigl[ 
\frac{a_0}{\ell^3}\frac{\media{\hat n_+^2}}\e
+\frac{a_0 n}{\ell^3}
+\e\media{\hat n_+}\frac{a_0 n}{\ell^3}\nonumber\\
&& \qquad \qquad
+\frac{a_0}{\ell^3}|n-\r\ell^3|\media{\hat n_+}
+\frac{a_0\media{\hat n_+^2}}{\ell^3}
+\e \frac{a_0 n}{\ell^3}\media{\hat n_+}
+\frac1\e\frac{a_0\media{\hat n_+}}{R^3}
+\frac1\e\frac{a_0\media{\hat n_+^2}}{R^3}
\Bigr]\,. \lb{2.61}\eea
Now, if $0<\e<1$, 
using $n_+\leq C n Y^{\nu_0}$, (\ref{2.42}), and and (\ref{2.50a}), 
we get from the last inequality that 
\be \media{H^n_{\ell}} \ge -nI-Cn\r a_0\Bigl[\e Y^{\n_0}+Y^{\frac32\n_0}
+\frac1\e Y^{2\n_0-3b-3d}\Bigr]\;.
\lb{2.62}\ee
Optimizing over $\e$ yields
\be \media{H^n_{\ell}}\ge -nI-Cn\r a_0\,Y^{\a_1}\;,\lb{2.63}\ee 
where $\a_1=\frac32\n_0-\frac32b-\frac32d =\frac34-\frac92b-6d$. 
If 
\be\frac34-\frac92b-6d>\frac12\;,\label{2.63a}\ee
the error term $n\r a_0
Y^{\a_1}$ is much smaller than $n\r a_0 Y^{1/2}$ and, therefore, 
\be E^n_\ell\ge -nI-n\r a_0\,o(Y^{1/2})\;.\lb{2.64}\ee

We are left
with estimating the constant $I$ defined by (\ref{2.43}).
It is not difficult to see that, under the 
assumptions made so far on the scaling exponents, 
\be I=4\p\r a_0\left(\frac{a_1}{a_0}-\frac{128}{15\sqrt\p}\sqrt{\r a_0^3}
+o\big(\sqrt{\r a_0^3}\big)\right)\;.\lb{2.76}\ee
%
The conditions (\ref{2.27b}), (\ref{2.46ab}) and 
(\ref{2.63a}) on the scaling exponents 
that we required for the proof to work can be summarized into the following
conditions:
\be 2b+d>6\t\;,\qquad \t>d\;,\qquad \frac16>3b+4d\;.\label{2.65}\ee
It is easy to check that if 
$d<1/69$ then all these requirements on the scaling exponents
can be satisfied. \qed


\begin{thebibliography}{38}
\expandafter\ifx\csname bibnamefont\endcsname\relax
  \def\bibnamefont#1{#1}\fi
\expandafter\ifx\csname bibfnamefont\endcsname\relax
  \def\bibfnamefont#1{#1}\fi
\expandafter\ifx\csname citenamefont\endcsname\relax
  \def\citenamefont#1{#1}\fi
\providecommand{\bibinfo}[2]{#2}

\bibitem[{\citenamefont{Bogololiubov}(1947)
\citenamefont{Bogoliubov}}]{Bo}
\bibinfo{author}{\bibfnamefont{N.~N.}~\bibnamefont{Bogoliubov}}:
\emph{\bibinfo{title}{On the theory of superfluidity}}
  \bibinfo{journal}{Izv. Akad. Nauk. USSR} \textbf{\bibinfo{volume}{11}},
  \bibinfo{pages}{77} (\bibinfo{year}{1947}). Engl. Transl. 
\bibinfo{journal}{J. Phys. (USSR)} \textbf{\bibinfo{volume}{11}},
  \bibinfo{pages}{23} (\bibinfo{year}{1947}).

\bibitem{LHY} T.D. Lee and C.N. Yang, {\it Many-Body Problem in
    Quantum Mechanics and Quantum Statistical Mechanics},
  Phys. Rev. {\bf 105}, 1119--1120 (1957).  T.D. Lee, K. Huang, and
  C.N. Yang, {\it Eigenvalues and Eigenfunctions of a Bose System of
    Hard Spheres and Its Low-Temperature Properties}, Phys. Rev. {\bf
    106}, 1135--1145 (1957).

\bibitem{ESY} L. Erd\"os, B. Schlein, and H.T. Yau, {\it The ground
    state energy of a low density Bose gas: a second order upper
    bound}, preprint arXiv:0806.4873

\bibitem[{\citenamefont{Conlon, Lieb and Yau}(1988)
\citenamefont{Conlon, Lieb, and Yau}}]{CLY}
\bibinfo{author}{\bibfnamefont{J.}~\bibnamefont{Conlon}},
  \bibinfo{author}{\bibfnamefont{E.~H.}~\bibnamefont{Lieb}}
 \bibnamefont{and} 
  \bibinfo{author}{\bibfnamefont{H.-T.}~\bibnamefont{Yau}}:
\emph{\bibinfo{title}{The $N^{7/5}$ Law for Charged Bosons}}
  \bibinfo{journal}{Commun. Math. Phys.} \textbf{\bibinfo{volume}{116}},
  \bibinfo{pages}{417-448} (\bibinfo{year}{1988}).

\bibitem[{\citenamefont{Girardeau and Arnowitt}(1959)
\citenamefont{Girardeau, and Arnowitt}}]{GA}
\bibinfo{author}{\bibfnamefont{M.}~\bibnamefont{Girardeau}}
\bibnamefont{and} 
  \bibinfo{author}{\bibfnamefont{R.}~\bibnamefont{Arnowitt}}:
\emph{\bibinfo{title}{Theory of Many-Boson Systems: Pair Theory}}
  \bibinfo{journal}{Phys. Rev.} \textbf{\bibinfo{volume}{113}},
  \bibinfo{pages}{755-761} (\bibinfo{year}{1959}).

\bibitem[{\citenamefont{Lieb and Solovej}(2001)
\citenamefont{Lieb, and Solovej}}]{LS}
  \bibinfo{author}{\bibfnamefont{E.~H.}~\bibnamefont{Lieb}}
 \bibnamefont{and} 
  \bibinfo{author}{\bibfnamefont{J.~P.}~\bibnamefont{Solovej}}:
\emph{\bibinfo{title}{Ground State Energy of the One-Component 
Charged Bose Gas}}
  \bibinfo{journal}{Commun. Math. Phys.} \textbf{\bibinfo{volume}{217}},
  \bibinfo{pages}{127-163} (\bibinfo{year}{2001}). Errata:  
\bibinfo{journal}{Commun. Math. Phys.} \textbf{\bibinfo{volume}{225}}, 
 \bibinfo{pages}{219-221} (\bibinfo{year}{2002}).

\bibitem[{\citenamefont{Lieb, Seiringer, Solovej and Yngvason}(2005)
\citenamefont{Lieb, Seiringer, Solovej, and Yngvason}}]{LSSY}
  \bibinfo{author}{\bibfnamefont{E.~H.}~\bibnamefont{Lieb}},
\bibinfo{author}{\bibfnamefont{R.}~\bibnamefont{Seiringer}},
  \bibinfo{author}{\bibfnamefont{J.~P.}~\bibnamefont{Solovej}}
 \bibnamefont{and} 
  \bibinfo{author}{\bibfnamefont{J.}~\bibnamefont{Yngvason}}:
\emph{\bibinfo{title}{The mathematics of the Bose gas and its condensation}},
Oberwolfach Seminars, \textbf{\bibinfo{volume}{34}},
Birkh\"auser Verlag, Basel (\bibinfo{year}{2005}).

\end{thebibliography}
\end{document}